\documentclass[letterpaper, 10 pt, journal, twoside]{IEEEtran}

\usepackage{url}
\usepackage{amsmath} % assumes amsmath package installed
\usepackage{amssymb}  % assumes amsmath package installed
\usepackage{amsthm}
\usepackage{lipsum,adjustbox}
\usepackage{tikz}
\usepackage{tikz-qtree}
\usepackage[font=footnotesize]{caption}
\usepackage{multirow}

\usepackage[shortlabels]{enumitem}

\usepackage{cite}
\usepackage{dsfont}

\DeclareMathOperator{\lie}{\mathcal{L}}
\DeclareMathOperator{\Lap}{\mathsf{L}}
\DeclareMathOperator{\N}{\mathcal{N}}
\DeclareMathOperator{\X}{\mathcal{X}}
\DeclareMathOperator{\A}{\mathsf{A}}

\DeclareMathOperator{\V}{\mathcal{V}}
\DeclareMathOperator{\E}{\mathcal{E}}
\DeclareMathOperator{\G}{\mathcal{G}}
\DeclareMathOperator{\Sc}{\mathcal{S}}
\DeclareMathOperator{\HH}{\mathcal{H}}

\newcommand{\uu}{\mathbf{u}}

\newcommand{\real}{\mathbb{R}}
\newcommand{\realpos}{\mathbb{R}_{>0}}
\newcommand{\realnn}{\mathbb{R}_{\geq 0}}
\newcommand{\integer}{\mathbb{Z}}
\newcommand{\integerpos}{\mathbb{Z}_{>0}}

\newcommand{\diag}{\operatorname{diag}}

\DeclareMathOperator{\zero}{\mathbf{0}}
\DeclareMathOperator{\xx}{\mathbf{x}}

\newtheorem{theorem}{Theorem}[section]
\newtheorem{proposition}[theorem]{Proposition}

\newtheorem{corollary}[theorem]{Corollary}
\theoremstyle{remark}
\newtheorem{remark}{Remark}
\theoremstyle{definition}

\newcommand{\longthmtitle}[1]{\mbox{} \emph{(#1):}}
\newcommand{\until}[1]{\{1,\dots,#1\}}
\newcommand{\setdef}[2]{\{#1 \; | \; #2\}}

\newcommand{\oprocendsymbol}{\hbox{$\bullet$}}
\newcommand{\oprocend}{\relax\ifmmode\else\unskip\hfill\fi\oprocendsymbol}

\allowdisplaybreaks

\title{Learning Invariant Stabilizing Controllers for Frequency Regulation under Variable Inertia
  \thanks{This work was supported by NSF
Award ECCS-1947050.}
}

\author{Priyank Srivastava  \quad Patricia Hidalgo-Gonzalez \quad Jorge
  Cort\'{e}s% <-this % stops a space}
  \thanks{P. Srivastava is with the Department of Mechanical Engineering, Massachusetts Institute of Technology, {\tt\small psrivast@mit.edu}
  
  P. Hidalgo-Gonzalez and J. Cort\'es are with the Department of Mechanical and
    Aerospace Engineering, UC San Diego,
    {\tt\small \{phidalgogonzalez, cortes\}@ucsd.edu}}%
}

\begin{document}

\maketitle
\thispagestyle{empty}
\pagestyle{empty}

\begin{abstract}
Declines in cost and concerns about the environmental impact of traditional generation have boosted the penetration of renewables and non-conventional distributed energy resources into the power grid. The intermittent availability of these resources causes the inertia of the power system to vary over time. As a result, there is a need to go beyond  traditional controllers designed to regulate frequency under the assumption of invariant dynamics. This paper presents a learning-based framework for the design of stable controllers based on imitating datasets obtained from linear-quadratic regulator (LQR) formulations for different  switching sequences of inertia modes. The proposed controller is linear and invariant, thereby interpretable, does not require the knowledge of the current operating mode, and is guaranteed to stabilize the switching power dynamics. We also show that it is always possible to stabilize the switched system using a communication-free local controller, whose implementation only requires each node to use its own state. Simulations on a  12-bus 3-region network illustrate our results. 
\end{abstract}

\section{Introduction}\label{sec:intro}
In power networks, any mismatch between electricity generation and consumption leads to the deviation of the frequency from its nominal value. The increasing penetration of renewable energy resources (RES), along with their intermittent availability, has made ensuring frequency regulation more relevant than ever. The presence of RES reduces the inertia of the system and makes it time-varying. As such, traditional controllers designed for invariant systems are no longer guaranteed to be stabilizing. Motivated by these considerations, this paper addresses the problem of optimally stabilizing the frequency of power networks with time-varying inertia.

\emph{Literature Review:}
In the traditional paradigm of power systems, there exists a number of mechanisms to prevent frequency excursions, cf.~\cite{PK:94,FD-JWSP-FB:16-tcns}. Inertial response is the first (automatic) response when any power imbalances occur.
It originates from the kinetic energy stored in synchronous generators and determines the instantaneous frequency when power imbalances arise. 
More inertia in the system translates into a slower rate of change of frequency.  As the frequency starts deviating, some generators respond proportionally to this deviation through the governor response or droop control~\cite{EE-MM-BK:11}. 
After droop control starts actuating, slower mechanisms (e.g., spinning reserves) participate to restore the frequency. 
RES, such as wind and solar, are usually connected through inverters, decoupling their rotational inertia (if existing) from the grid. 
As a result, the system inertia is an inverse function of the number of RES.
In fact, since different distributed energy resources make autonomous decisions when connecting to the grid, the inertia of the system becomes time-varying~\cite{AU-TSB-GA:14}. 
This can provoke abrupt variation in the grid frequency under mismatches of generation and demand. Without appropriate measures, this can make the standard frequency control schemes too slow to mitigate arising contingencies. The impact of low inertia in the future grid is captured by system operators in various reports~\cite{AEMO:16,ENTSO-E:16,ERCOT:13}.

%In the traditional paradigm of power systems, where generation has been dominated by synchronous generation, the inertia has been considered constant. However, in recent years, it has been observed that due to the increase in generation from RES, the rotational inertia in the network has become lower and time-varying \cite{AU-TSB-GA:14, ERCOT}. 
%

A growing body of work addresses this need by analyzing the effect of inertia variations on frequency~\cite{TSB-TL-DJH:15}, designing robust controllers~\cite{GSM-SC-TW:18}, and identifying conditions on the power supply dynamics and rate of change of inertia that ensure stability~\cite{AK-ST-MP:21}.
%The work~\cite{PHG-DSC-RD-RHA-CJT:18} proposes a switched affine hybrid system framework~\cite{DL:03} to model power dynamics taking into account inertia variability. To stabilize such systems,~\cite{PHG-RHA-DSC-CJT:19}
The work~\cite{PHG-RHA-DSC-CJT:19} uses a switched affine hybrid system framework  to model inertia variations and proposes a learning-based invariant controller stabilizing each inertia mode of the closed-loop dynamics.
%where each mode of the closed-loop switched system is stabilized. %a posteriori of the training phase.
%This is not sufficient to conclude the stability of the switched system and in general, could lead to instability for some switching sequences.
This formulation is extended in~\cite{PHG-RHA-DSC-CJT:19b} to 
enhance sparsity, albeit each node needs to communicate with a certain minimum threshold number of nodes,  and take into account the stability of the switched system, albeit there is no guarantee that a feasible solution exists. Both~\cite{PHG-RHA-DSC-CJT:19,PHG-RHA-DSC-CJT:19b} assume that all the nodes have equal inertia in each mode and stability is considered a posteriori once the training is complete.
Ideally, as pointed out in~\cite{RD-PHG-SK-RHA-GH-DSC:20}, stability guarantees should be encoded in the training phase itself. In fact, the lack of guarantees on stability is a shortcoming in many works employing machine learning techniques for power systems, 
%to leverage optimality of the control action at the same time as stability
%
%\marginJC{Safety? Also, the reference to [10] is to mean that such embedding is done there? It's unclear.}
%
% Most of the existing work employing machine learning techniques for power systems does not take safety guarantees into account 
cf.~\cite{WW-NY-JS-YG:19,DY-MZ-DS:11,MHK-MG:21,MG-RF-DE:17}. 
For example, the work in~\cite{MG-RF-DE:17} presents an overview on reinforcement learning (RL) techniques for power systems,
% discussing general trends, common algorithms used, and specific power system applications. However, it does not 
but does not touch upon the stability aspects. The work~\cite{DE-MG-LW:04} discusses the importance of stability when using RL in power systems and how key RL assumptions may not hold in some power systems applications. 
%More recently, two contributions have been made at the intersection of control theory and ML in power systems \cite{WC-BZ:20,MJ-JL:20}. These works have developed RL approaches with stability guarantees.
Recently,~\cite{WC-BZ:20,MJ-JL:20} have developed RL approaches for frequency control with stability guarantees, but the designed controllers do not consider time-varying frequency dynamics due to the changing inertia.

%However, RL as an online technique, places this approach as a different method compared to the work previously discussed (offline training) opening a different set of challenges and implementation considerations.

\emph{Statement of Contributions:}
We consider the problem of designing an invariant controller to
stabilize the frequency of a power network with time-varying
inertia. The fact that the controller is invariant makes it oblivious
to changes in inertia, hence facilitating its implementability by
power system operators.
Our starting point is a formulation of the frequency dynamics of the
power network as a switched affine system, where each mode corresponds
to a different value of the inertia.
% 
%\marginJC{Why is this not in the lit review: "Compared to the
%existing works in literature using this formulation, we do not assume
%the inertia of all the nodes to be same."?}  
%
To address the fact that changes in the operating mode are not known a
priori, we consider a candidate set of switching sequences and, for
each of them, solve a finite-horizon LQR problem to generate optimal
state-input trajectories to be used as data.  We then formulate the
controller design problem as a constrained least-squares optimization,
where the objective function measures the fit of the trajectories
generated with the controller to the data, and the constraints encode
the stabilization requirement for the switched system.  Our first
result considers the formulation where constraints correspond to the
stabilization of the individual modes and its proof is constructive,
providing an explicit stabilizing controller which is distributed over
the power network.  Our second result generalizes our treatment to
guarantee the system stability under arbitrary switching, and
establishes that regardless of the inertia of the operating mode,
stabilization is always possible using an invariant controller. Our
last result shows that, in fact, there always exists a stabilizing
controller which is local, meaning that its implementation only
requires each node to use its own state. Simulations on a 12-bus
3-region network with variable inertia demonstrate the stabilizing
performance of the learned controllers with and without sparsity
constraint.

\section{Problem Formulation}
%\subsection{Time-Varying Power Dynamics}
Consider\footnote{Throughout the paper, we use the following notation. Let $\real, \realnn, \realpos, \integer, \integerpos$ denote the set of reals, non-negative reals, positive reals, integers, 
and positive integers, respectively. We denote by $|\X|$ the cardinality of a set~$\X$. The symbol
$\zero$ represents the matrix of all zeros and $I$ denotes the identity matrix, with appropriate dimensions. For a matrix $A$, 
$A_{ij}$ denotes its $ij$th element,
$A^\top$ denotes its transpose, and $A^{-1}$ its inverse.
$A \succ \zero$ $(\succeq \zero)$ and $A \prec \zero$ $(\preceq \zero)$ denote, respectively, that $A$ is positive definite (semidefinite) and negative definite (semidefinite). $A \otimes B$ denotes the Kronecker product of  matrices $A$ and $B$.  $\diag(a_i)$ is the matrix with entries $\{a_i\}_{i=1}^m$ in its main diagonal.
With a slight abuse of notation, we let $(x;y) \in \real^{m+n}$ denote the concatenated vector obtained by putting together the entries of vectors $x \in \real^m$ and $y \in \real^n$. We also employ basic concepts from graph theory following~\cite{CDG-GFR:01}. We denote a weighted undirected graph by $\G = (\V , \E , \A)$, with $\V$ as the set of nodes and $\E \subseteq \V \times \V$ as the set of edges. $(i,j) \in \E$ iff $(j,i) \in \E$  iff there is an edge from node $i$ to $j$. 
A node $j \in \V$ is a neighbor of $i$ if $(i,j) \in \E$.
We denote the set of neighbors of node $i$ by $\N_i$.
With $|\V| = n$, the \emph{adjacency matrix} $\A \in \real^{n \times n}$ of $\G$ is such that $\A_{ij} > 0$ if $(i,j) \in \E$ and $\A_{ij} =0$, otherwise. The weighted degree of node $i$ is $d(i)=\sum_{\N_i} \A_{ij}$. Finally, the \emph{Laplacian matrix} $\Lap \in \real^{n \times n}$ is~$\Lap=\diag(d(i))-\A$.
} a power network with $n \in \integerpos$ nodes, whose interconnection is described by an undirected graph $\G$.
Following~\cite{BKP-SB-FD:17}, we consider a DC approximation of the power flow. The frequency and phase angle dynamics for each node $i \in \until{n}$ are approximated as follows
\begin{align*}%\label{eq:swing}
    m_i \ddot{\theta}_i + d_i \dot{\theta}_i = u_i - \sum\limits_{\N_i} b_{ij} (\theta_i - \theta_j) ,
\end{align*}
where $u_i$ is the power input at node $i$ and $b_{ij} \in \realnn$ is the susceptance between lines $i$ and $j$.
 If node $i$ is a synchronous generator, then $\theta_i \in \real$ denotes the rotor angle, $m_i \in \realpos$ the rotational inertia of the generator $i$ and $d_i \in \realpos$ the primary speed droop control at node $i$.
 If node $i$ corresponds to a renewable or battery interfaced via a power electronics converter, then $\theta_i$ is the voltage phase angle, $m_i$ is the power measurement time constant or the virtual inertia through a controlled device, and $d_i$ is the droop control coefficient. The joint
state-space representation of the network
is
\begin{align}\label{eq:swing_invariant}
\begin{bmatrix}
\dot{\theta} \\ \dot{\omega} \end{bmatrix}=
\underbrace{\begin{bmatrix} \zero & I \\ -M^{-1} \Lap & -M^{-1}D \end{bmatrix}}_{A} \begin{bmatrix} \theta \\ \omega\end{bmatrix}+ 
\underbrace{\begin{bmatrix} \zero \\ M^{-1}\end{bmatrix}}_{B} u,
\end{align}
where $x=(\theta;\omega) \in \real^{2n}$ corresponds to the stacked vector of angle and frequency deviations at each node,
$M = \diag(m_i) \in \real^{n \times n}$ is the diagonal matrix with inertia coefficients, 
$D = \diag(d_i) \in \real^{n \times n}$ is the diagonal matrix with droop control coefficients, and
$\Lap$ is the Laplacian of the weighted version of $\G$ whose adjacency matrix is $\A_{ij}=b_{ij}$, $i,j \in \until{n}$.
One can verify that $(A, B)$ is stabilizable.
%, and 
%$u \in \real^n$ corresponds to the vector of power inputs.

The formulation~\eqref{eq:swing_invariant} assumes that the inertia of the system remains constant and makes sense in the traditional paradigm of power systems. However, in scenarios with increasing penetration of renewables, the inertia of the network may change over time. Hence, it is reasonable to incorporate the time dependence in the inertia at each node. We do this by considering a switched-affine system representation as in~\cite{PHG-DSC-RD-RHA-CJT:18}, where each mode corresponds to a different value of the inertia. If  $m \in \integerpos$ is the number of modes, the frequency dynamics are then given by
\begin{align}\label{eq:swing_modes}
\begin{bmatrix}
\dot{\theta} \\ \dot{\omega} \end{bmatrix}=
\underbrace{\begin{bmatrix} \zero & I \\ -M_{q(t)}^{-1} \Lap & -M_{q(t)}^{-1}D \end{bmatrix} }_{A_{q(t)}} 
\begin{bmatrix} \theta \\ \omega\end{bmatrix}+ 
\underbrace{\begin{bmatrix} \zero \\ M_{q(t)}^{-1}\end{bmatrix} }_{B_{q(t)}} u.
\end{align}
Here, at time $t$, the system is in mode $q(t) \in \until{m}$ and $M_q(t)$ denotes the inertia of the network in mode $q(t)$.
The inertia at time $t$ depends on the online generators and the connected power electronics converters at that time. When convenient, we drop the argument $t$ and refer to $q(t)$ as~$q$.

Our goal is to design an optimal controller that brings the system~\eqref{eq:swing_modes} to the origin from any initialization.
%
%\marginJC{Where are the perturbations modeled? I don't see anything in the equations. Looks like we're just stabilizing. Question: if a switched linear system is stabilized with a common controller, as we do eventually, is convergence exponential? Is the switched-system ISS? If it is, then what we say here would make sense.}
%
Since we might not have knowledge of the current operating mode at all times, our aim is to design a time-invariant controller of the form
\begin{align*}
    u=K x,
\end{align*}
that stabilizes~\eqref{eq:swing_modes}, minimizes the state deviation, and optimizes the control input required. For a fixed linear system, this is  achievable using the solution to the linear-quadratic control (LQR) problem. However, for the switched system, this cannot be done unless the switching sequence is known beforehand. Optimizing instead for all possible switching sequences quickly becomes computationally intractable. Therefore, we follow an offline, data-driven, imitation-based approach that balances the goals of optimality and stability: the basic idea is to consider a set of candidate switching sequences, solve a finite-horizon LQR problem for each of them, and finally use the resulting trajectories as a training set to design a stabilizing controller imitating the observed behavior.

\section{Data-Driven Controller Design}\label{sec:controller}
In this section, we carry out our approach to design a common stabilizing time-invariant controller using training data 
generated for system~\eqref{eq:swing_modes} for a variety of scenarios.
We start by describing  in Section~\ref{sec:train_data} how the data is generated via a finite-horizon LQR formulation.
Then we provide in Section~\ref{sec:simultaneous} a least-squares formulation to learn the controller while guaranteeing the stability of each mode $q \in \until{m}$. Since the stability of all the modes is not sufficient to guarantee the stability of the switched system, we generalize in Section~\ref{sec:switched} our treatment to the stabilization of the switched system via a common Lyapunov function. 

\subsection{Training Data from Optimal Input Trajectories}\label{sec:train_data}
In order to generate the training data which would later be used to learn the controller gain $K$, we solve $\Sc \in \integerpos$ instances of the finite-horizon LQR problem
\begin{align}\label{eq:training}
& \min_{\xx,\uu} & &  \int\limits_{0}^T \big( x(t)^\top Q x(t) + u(t)^\top R u(t) \big) \; dt  \\
& \text{s.t.} & & x(0)=x_0 \nonumber \\
&&& \dot{x}(t)=A_{q(t)}x(t)+ B_{q(t)} u(t), \quad t \in [0,T] \nonumber,
\end{align}
where $Q \succeq \zero \in \real^{2n \times 2n}$ penalizes state deviations, $R \succ \zero \in \real^{n \times n}$ represents a cost associated to the control action, $T>0$ is the time horizon, $x_0 \in \real^{2n}$ is the initial state, and $\xx(t) \in \real^{2n}$ and $\uu(t) \in \real^{n}$ are the variables describing the optimal state and input trajectories, respectively. 

We generate $\Sc$ scenarios by selecting different initial conditions $x_0$ and switching sequences $q(t)$, with the pair $(\xx^k(t), \uu^k(t))$ denoting the training data for scenario $k \in \until{\Sc}$. The scenarios provide data in the form of desirable trajectories for the controller to imitate. 
The amount of information available to capture optimality grows with the number of scenarios considered, at the cost of an increasing computational effort to handle them. 
Also, the number of trajectories by itself does not guarantee that the resulting controller is stable. Instead, in our design formulations below, we make sure the stability of the controller is guaranteed independently of the number of scenarios considered. Regarding the selection of initial conditions for the scenarios, since the frequency deviation is usually bounded for real systems, from a practical viewpoint, rather than taking  them to be uniformly distributed throughout the state space, it makes sense to consider initial conditions close to the origin.

\subsection{Simultaneous Stabilization of All Switching Modes}\label{sec:simultaneous}
Here, we are interested in designing a learned time-invariant controller which guarantees stability for each mode $q \in \until{m}$.
Let $\HH$ denote the set of Hurwitz matrices.
Then the controller design problem described above can be cast as an optimization of the form 
\begin{align}
& \min_{K}  & & \sum\limits_{k=1}^{\Sc} \int\limits_{0}^T \|\uu^k(t) - K \xx^k(t) \|_2^2 \; dt \label{eq:simultaneous} 
\\
& \text{s.t.} & &  A_{q} + B_{q} K \in \HH,  \quad \forall q 
%\in \until{m} 
\nonumber.
\end{align}
Since the set of Hurwitz matrices is not convex,~\eqref{eq:simultaneous} is non-convex.
In fact, finding a feasible solution of~\eqref{eq:simultaneous}, also referred to as the \emph{simultaneous stabilization} problem, is NP-hard for general system and input matrices, cf.~\cite{VB-JNT:97}.
However, the matrices $\{A_{q}\}_{q=1}^m$ and $\{B_{q}\}_{q=1}^m$ in our setup are not arbitrary,
and indeed have a well-defined structure. Specifically, the only quantity that specifies the operating mode $q \in \until{m}$ is the inertia matrix $M_{q}$. Building on this insight, we prove that the simultaneous stabilization problem~\eqref{eq:simultaneous} is always feasible.
Our proof is constructive and relies on identifying a controller that stabilizes all the modes.

\begin{proposition}
\longthmtitle{Feasibility of the simultaneous stabilization data-driven problem for individual modes}\label{prop:simultaneous}
Problem~\eqref{eq:simultaneous} is always feasible.
\end{proposition}
\begin{IEEEproof}
Let $K=[K_1 \quad  K_2]$, where $K_1, K_2 \in \real^{n \times n}$. Then from equation~\eqref{eq:swing_modes}, the closed-loop system matrix for mode $q \in \until{m}$ is given by
\begin{align}
A_q + B_q K &= \begin{bmatrix} \zero & I \\ -M_q^{-1} \Lap & -M_q^{-1} D \end{bmatrix}+
 \begin{bmatrix} \zero \\ M_q^{-1} \end{bmatrix} \begin{bmatrix} K_1 & K_2 \end{bmatrix} \nonumber \\
&=\begin{bmatrix} \zero & I \\ -M_q^{-1} (\Lap-K_1) & -M_q^{-1} (D-K_2) \end{bmatrix}. \label{eq:closed_loop}
\end{align}
Let us first consider the case when, in a given mode $q$, the inertia coefficient of all the nodes is the same, and is given by $m_q \in \realpos$.
Then we have $M_q=m_q I$. 
Choosing 
\begin{align}\label{eq:K}
K_1=\Lap-I \text{ and } K_2=D-I,
\end{align}
the closed-loop system matrix~\eqref{eq:closed_loop} becomes
\begin{align}
A_q + B_q K &= \begin{bmatrix} \zero & I \\ -1/m_q I & -1/m_q I \end{bmatrix} \nonumber \\ 
& = \underbrace{\begin{bmatrix} 0  & 1 \\ -1/m_q & -1/m_q \end{bmatrix}}_{S_q} \otimes I. \label{eq:closed_loop_same}
\end{align}
The eigenvalues of the $2 \times 2$ matrix $S_q$ are negative for all $m_q > 0$. Hence, $A_q + B_q K \in \HH$ for all $q \in \until{m}$ 
%with the selected controller
.

Next we consider the general case where each node $i \in \until{n}$ might have a different inertia coefficient. Once again, choose $K_1$ and $K_2$ according to~\eqref{eq:K}. The closed-loop system matrix~\eqref{eq:closed_loop} now takes the form
\begin{align*}
A_q + B_q K = \begin{bmatrix} \zero & I \\ -M_q^{-1}  & -M_q^{-1}  \end{bmatrix} %\label{eq:closed_loop_different}.
\end{align*}
For each mode $q \in \until{m}$, consider the Lyapunov function candidate  $V_q:\real^{2n} \to \real$
\begin{align*}
V_q=x^\top \underbrace{\begin{bmatrix} I & \zero \\ \zero & M_q \end{bmatrix}}_{P_q} x.
\end{align*}
The Lie derivative of $V_q$ is given by
\begin{align*}
\lie_f V (x) &= x^\top \big( (A_q + B_q K)^\top P_q + P_q (A_q + B_q K) \big) x 
\\
&= x^\top \begin{bmatrix} \zero & \zero \\ \zero & -2I\end{bmatrix} x \le 0.
\end{align*}
This means that each mode $q \in \until{m}$ is stable,
and the result follows.
\end{IEEEproof}

The proof of Proposition~\ref{prop:simultaneous} considers first the case of equal inertia at each node, and then generalizes the argument to the case of different inertia at the nodes. Although establishing the feasibility of the simultaneous stabilization problem~\eqref{eq:simultaneous} in the former case is a special case of the latter, it is interesting to consider it separately as the eigenvalues of the closed-loop system can be explicitly characterized.

\begin{remark}\longthmtitle{Distributed learned controller stabilizing all the modes}
The proof of Proposition~\ref{prop:simultaneous} is constructive and relies on identifying a (not necessarily optimal) controller stabilizing all the modes. It is interesting to note that the controller identified in~\eqref{eq:K} is distributed over $\G$, meaning that to implement it, each node $i \in \until{n}$ needs to know just its angle and frequency, and the angle of the nodes to which it is electrically connected. \oprocend
\end{remark}

\subsection{Simultaneous Stabilization of the Switched System}\label{sec:switched}
The controller resulting from the simultaneous stabilization problem~\eqref{eq:simultaneous} in Section~\ref{sec:simultaneous} guarantees the stability of each individual mode, but does not guarantee the stability of the overall switched system~\eqref{eq:swing_modes} in general, cf.~\cite{MSB:98}.
To address this, here we reformulate the synthesis of the learned time-invariant controller by specifying  a common Lyapunov function as a certificate of its correctness. Formally, the controller design problem takes now the form
\begin{align}
& \min_{K,P} & & \sum\limits_{k=1}^{\Sc} \int\limits_{0}^T \|\uu^k(t) - K \xx^k(t) \|_2^2 \; dt \label{eq:common_lyap}
\\
& \text{s.t.} & & (A_q + B_q K)^\top \! P \! + \! P (A_q + B_q K) \prec \zero, \; \forall q  
%\in \until{m} 
\nonumber\\
& & & P \succ \zero \nonumber.
\end{align}
In this formulation, we aim to find a common quadratic Lyapunov function given by $V(x) = x^\top P x$. The first constraint in~\eqref{eq:common_lyap} ensures that the Lie derivative of the Lyapunov function along the evolution of~\eqref{eq:swing_modes} remains negative for each mode $q \in \until{m}$, thereby guaranteeing the stability of the switched system.
Note that the problem~\eqref{eq:common_lyap} is bilinear in the decision variables $K$ and $P$ and, hence, nonconvex.
The next result establishes the feasibility of problem~\eqref{eq:common_lyap}.
% Reformulated Problem
%  \begin{align}
% & \min_{Y,X} & & \sum\limits_{k=1}^K \sum\limits_{t=1}^T \|u_t^k - YX^{-1} x_t^k \|_2^2 \label{eq:common_lyap_re} \\
% & \text{s.t.} & & A_q X + X A_q^\top + B_q Y + Y^\top B_q^\top  \prec \zero \quad \forall q \nonumber\\
% & & & X \succ \zero \nonumber.
% \end{align}
% The controller $K$ is then given by
% \begin{align}
% K=Y X^{-1}.
% \end{align}
% \textbf{Same inertia}

% Let $\overline{m}=\max m_q$.
% and $\underline{m}=\min m_q$.
\begin{theorem}\longthmtitle{Feasibility of the  simultaneous stabilization data-driven problem for the switched system}\label{thm:switched}
Problem~\eqref{eq:common_lyap} is always feasible.
\end{theorem}
\begin{IEEEproof}
Using $X=P^{-1}$ and $K=YX^{-1}$, cf.~\cite{GED-FP:00}, the constraints in problem~\eqref{eq:common_lyap} can be equivalently written as
\begin{subequations}\label{eq:common_lyap_re}
\begin{align}
& A_q X \!+\! X A_q^\top \!+\! B_q Y \!+\! Y^\top B_q^\top  \prec \zero, \; \forall q 
%\in \until{m}  
\label{eq:common_lyap_re_a}\\
& X \succ \zero . \label{eq:common_lyap_re_b}
\end{align}
\end{subequations}
With $X_1, X_2, X_3, Y_1, Y_2 \in \real^{n \times n}$, let $X=\begin{bmatrix}
X_1 & X_2 \\ X_2^\top & X_3
\end{bmatrix}$ and $Y= \begin{bmatrix} Y_1 & Y_2 \end{bmatrix}$.
Then using the structure of $\{A_q\}_{q=1}^m$ and $\{B_q\}_{q=1}^m$, constraint~\eqref{eq:common_lyap_re_a} can be rewritten as
\begin{align*}
&\begin{bmatrix} \zero & I \\ -M_{q}^{-1} \Lap & -M_{q}^{-1}D \end{bmatrix}
\begin{bmatrix} X_1 & X_2 \\ X_2^\top & X_3
\end{bmatrix} + \begin{bmatrix} \zero  \\ M_{q}^{-1}  \end{bmatrix}
\begin{bmatrix} Y_1 & Y_2 \end{bmatrix}
+ \\ &\begin{bmatrix} X_1 & X_2 \\ X_2^\top & X_3
\end{bmatrix} \begin{bmatrix} \zero & -\Lap M_{q}^{-1} \\ I & -D M_{q}^{-1}\end{bmatrix} 
+\begin{bmatrix} Y_1^\top \\ Y_2^\top \end{bmatrix} 
\begin{bmatrix} \zero &  M_{q}^{-1}  \end{bmatrix}
\prec \zero , 
\end{align*}
for all $ q \in \until{m}$. Performing the matrix multiplications and using the abbreviated notation 
\begin{align*}
  Z_q&=-M_q^{-1} \Lap X_1 - M_q^{-1} D X_2^\top + M_q^{-1} Y_1
  \\
  W_q&=M_q^{-1} \Lap X_2 + M_q^{-1} D X_3,
\end{align*}
the inequality can be further rewritten as
\begin{align*}
    \begin{bmatrix}
    X_2^\top & X_3 \\ Z_q & -W_q + M_q^{-1} Y_2
    \end{bmatrix} + \begin{bmatrix}
    X_2 & Z_q^\top \\ X_3^\top & -W_q^\top + Y_2^\top M_q^{-1} 
    \end{bmatrix} \prec \zero ,
\end{align*}
for all $ q \in \until{m}$. 
Hence,~\eqref{eq:common_lyap_re_a} is satisfied if the matrix
\begin{align*}%\label{eq:matrix_nd}
   & \begin{bmatrix} -X_2 - X_2^\top & -X_3-Z_q^\top \\
-X_3^\top-Z_q & W_q + W_q^\top - M_q^{-1}Y_2 - Y_2^\top M_q^{-1} \end{bmatrix}
\end{align*}
is positive definite for all $ q \in \until{m}$. 
Using the Schur complement, cf.~\cite{SB-LV:04}, the positive definiteness condition (and hence~\eqref{eq:common_lyap_re_a}) is equivalent to
\begin{subequations}\label{eq:schur}
\begin{align}
-X_2-X_2^\top \succ &\zero, \label{eq:schur_a}
\\
W_q+W_q^\top -M_q^{-1}Y_2 - Y_2^\top M_q^{-1} \succ &\zero, \label{eq:schur_b} 
\\
W_q+W_q^\top -M_q^{-1}Y_2 - Y_2^\top M_q^{-1} + \qquad & \notag 
\\ (X_3^\top + Z_q) (X_2 + X_2^\top)^{-1} (X_3 + Z_q^\top) \succ &\zero, \label{eq:schur_c}
\end{align}
\end{subequations}
for all $ q \in \until{m}$. Now, 
choose $X \succ \zero$ satisfying~\eqref{eq:schur_a}.
Then, since $W_q$ and $Z_q$ are independent of $Y_2$, there exists $N \prec \zero \in \real^{n \times n}$, independent of $Y_2$ too, such that 
%$N \prec \zero$, and 
for all $q \in \until{m}$,
\begin{subequations}\label{eq:N}
\begin{align}
W_q+W_q^\top - N \succ & \zero , 
\\
(X_3^\top + Z_q)(X_2+X_2^\top)^{-1} (X_3 + Z_q^\top)  - N  \succ & \zero.
\end{align}
\end{subequations}
Finally, using~\eqref{eq:schur_b}-\eqref{eq:schur_c} and~\eqref{eq:N} along with the fact that $N \prec \zero$, it suffices to show  that there exists $Y_2$ such that $ 2N - M_q^{-1}Y_2 - Y_2^\top M_q^{-1} \succ \zero$, for all~$q \in \until{m}$. Let $\overline{M}$ denote the matrix obtained after taking the entry wise maximum of the inertia coefficient matrix at all nodes. Then the above inequality is satisfied if $Y_2$ is chosen such that
\begin{align}\label{eq:Y_2}
Y_2 \prec N\overline{M},
\end{align}
completing the proof.
\end{IEEEproof}
 
Proposition~\ref{prop:simultaneous} can be considered as a special case of Theorem~\ref{thm:switched}. However, the results differ in their proof methodologies. The proof of Proposition~\ref{prop:simultaneous} provides an explicit expression for a feasible controller, which in addition is distributed over $\G$.
This, however, does not mean that the optimizer of~\eqref{eq:simultaneous} is distributed (although it does imply that one can look for solutions of~\eqref{eq:simultaneous} among controllers that are distributed over $\G$). Instead, the proof of Theorem~\ref{thm:switched} identifies an ordered sequence of steps that lead  to the identification of a controller stabilizing the switched system. In principle, 
there is no guarantee that the resulting controller will be distributed. %Nevertheless, Theorem~\ref{thm:switched} serves as the basis for the design of a distributed controller in the next section.
% \subsection{Simultaneous Stabilization via Distributed Control}\label{sec:local_control}
The  following result shows that a distributed controller does in fact exist. More precisely, it shows that there exists a  controller that does not need communication even with  neighboring nodes (we term this special form of distributed controller as \emph{local}). The proof methodology leverages the freedom in choosing various parameters in the proof of Theorem~\ref{thm:switched}. 

\begin{corollary}
 \longthmtitle{Local controller stabilizing the switched system}
 \label{co:local}
There exists a controller of the form $u=D_1 \theta + D_2 \omega$, 
where $D_1, D_2 \in \real^{n \times n}$ are diagonal matrices, satisfying the constraints in problem~\eqref{eq:common_lyap}.
\end{corollary}
 \begin{IEEEproof}
Following the proof of Theorem~\ref{thm:switched}, we are interested in identifying $X$ and $Y$ satisfying~\eqref{eq:common_lyap_re}. 
Let us choose $X_1=I,~X_2=-I$, and $Y_1=\zero$. Then using the Schur complement, \eqref{eq:common_lyap_re_b} is satisfied iff 
\begin{align}\label{eq:X_3}
X_3 - I \succ \zero.
\end{align}
To satisfy~\eqref{eq:common_lyap_re_a}, once again, following the same steps as in the proof of Theorem~\ref{thm:switched}, 
choose $N$ satisfying~\eqref{eq:N} and then,
$Y_2$ as a diagonal matrix satisfying~\eqref{eq:Y_2}.
%
%\marginJC{Why don't you write what $W$ and $Z$ are equal to in this case, so that you can also write (12) explicitly?}
%
The controller $K$ is then given by
$
    K = \begin{bmatrix}
    \zero & Y_2 
    \end{bmatrix}\begin{bmatrix}
    I & -I \\ -I & X_3
    \end{bmatrix}^{-1}
$.
Using the formula for the inverse of a partitioned matrix~\cite[Section 0.7.3]{RAH-CRJ:85},
\begin{align*}
    K= & \begin{bmatrix}
    \zero & Y_2
    \end{bmatrix} \begin{bmatrix}
    (I-X_3^{-1})^{-1} & (X_3-I)^{-1} \\
    (X_3-I)^{-1} & (X_3-I)^{-1}
    \end{bmatrix} \\
  =  & \begin{bmatrix}
    Y_2 (X_3-I)^{-1} & Y_2(X_3-I)^{-1}
    \end{bmatrix}.
\end{align*}
Now if one chooses $X_3$ to be a diagonal matrix (making the controller local) satisfying~\eqref{eq:Y_2} and~\eqref{eq:X_3},
then the resulting controller stabilizes the switched system.
\end{IEEEproof}
 
Although this result guarantees the existence of a local stabilizing controller, restricting the feasible set of~\eqref{eq:common_lyap} to controllers of that form could significantly affect the optimal value of the objective function. 
% In fact, compared to the stabilizing learned controllers in~\eqref{eq:common_lyap} and~\eqref{eq:sparse}, the associated cost could be larger by orders of magnitude, something that we have observed in simulations.
Motivated by Corollary~\ref{co:local} and this observation, we propose a middle ground that reformulates the optimization problem to promote sparsity in the learned controller. Formally, following~\cite{NKD-MRJ-ZQL:14}, let $\beta > 0$ be a design parameter that specifies the importance of promoting sparsity as compared to the original objective function of matching the data provided by the sampled optimal trajectories. 
Let $\E^c=\setdef{(i,j)}{(i,j) \notin \E}$, denote the set of indices whose corresponding vertices are not neighbors in $\G$.
The sparse-promotion controller design problem takes the form
\begin{align}
& \min_{K,P} & & \! \sum\limits_{k=1}^{\Sc} \int\limits_{0}^T \|\uu^k(t) - K \xx^k(t) \|_2^2 \; dt + \beta %\sum\limits_{\substack{i=1\\i \ne j}}^n 
\sum\limits_{(i,j) \in  \E^c} |K_{ij}| \label{eq:sparse} 
\\
& \text{s.t.} & & (A_q + B_q K)^\top \! P \! + \! P (A_q + B_q K) \prec \zero, \; \forall q  
%\in \until{m} 
\notag
\\
& & & P \succ \zero \notag.
\end{align}
Since~\eqref{eq:common_lyap} is feasible by Theorem~\ref{thm:switched}, problem~\eqref{eq:sparse} is feasible too.  To find a local controller, one could also consider a modified version of~\eqref{eq:sparse} where all the non-diagonal entries of $K$ are penalized.
% Note, however, that these are heuristic methods, and in general, there is no guarantee for the resulting learned controller to be distributed over $\G$ or local. 

\section{Simulations}\label{sec:sims}
\begin{figure}[htb]
    \centering
    \includegraphics[width=0.47\textwidth]{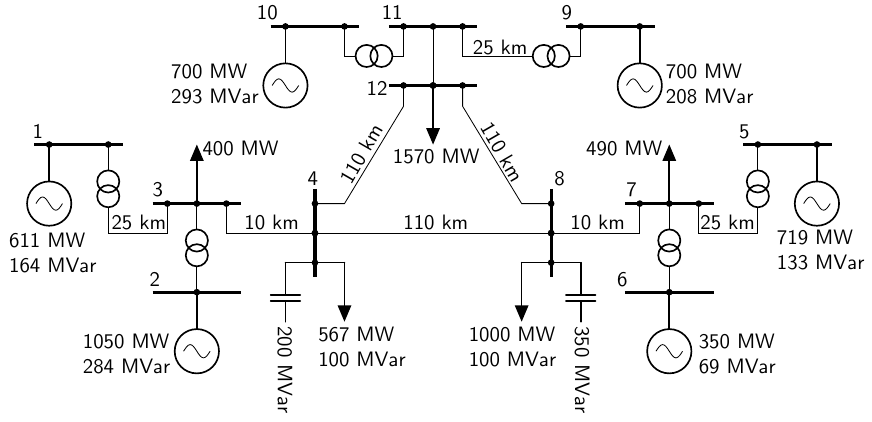}
    \caption{The 12-bus 3-region network used in simulations.}
    \label{fig:network}
\end{figure}

In this section, we demonstrate the effectiveness of the proposed approach via numerical experiments. We use the standard 12-bus 3-region network, shown in Figure~\ref{fig:network}, that has also been used in~\cite{PHG-DSC-RD-RHA-CJT:18,PK:94,BKP-SB-FD:17}.
We take $m=10$ and assume that at a given time $t$, the rotational inertia for each node $i \in \until{n}$ is same. Hence, each mode $q \in \until{m}$ of the hybrid system is given by one value of inertia in the set \{0.2, 0.5, 1, 1.5, 2, 2.5, 3, 3.5, 5, 9\}.
To generate the training data-set, we use $Q=\begin{bmatrix}
I & \zero \\ \zero & 10^5 I 
\end{bmatrix}$ and $R=10I$.
To implement~\eqref{eq:training}, we use its discrete-time counterpart
\begin{align}\label{eq:training_discrete}
& \min_{\xx,\uu} & &  \sum\limits_{t=0}^T x_t^\top Q x_t + u_t^\top R u_t  \\
& \text{s.t.} & & x_{t+1}=A^d_{q(t)}x_t+ B^d_{q(t)} u_t, \quad t \in \{0,\ldots,T-1\} \nonumber,
\end{align}
where $A^d_{q(t)} \real^{2n \times 2n}$ and $B^d_{q(t)} \in \real^{2n \times n}$ are respectively, the state and input matrices of the discretized system using a zero-order hold.
We use a stepsize of $10^{-2}$ seconds and simulate 50 scenarios of~\eqref{eq:training_discrete}, each for 50 time steps, using \verb|cvx|~\cite{MG-SB:14-cvx}. The initial conditions for all the scenarios (for both the angles and the frequencies) are different, and drawn from a normal distribution with 0 mean and 0.1 variance.
Each scenario starts in mode 7 (3 seconds of inertia), and from there, based on
a uniform distribution draw, the inertia of the
system can remain the same, increase, or decrease every 2 time steps.

We design three sets of controllers: 
(a) \emph{Optimal}: To design the first optimal and stable learned controller, %(referred as ``Optimal controller''),
we solve~\eqref{eq:common_lyap}
using the BMI algorithm in~\cite{QD-SG-WM-MD:12}. 
Since the algorithm requires a feasible initialization, we solve the feasibility problem associated with the LMI constraints~\eqref{eq:common_lyap_re} using \verb|cvx| to find an initial point. (b) \emph{Distributed}: To design the second learned controller, which is stable and sparse, %(referred to as ``Distributed controller''),
we solve~\eqref{eq:sparse} for various values of $\beta$, again using the algorithm in~\cite{QD-SG-WM-MD:12}. 
The controller turns out to be distributed over~$\G$ for $\beta=100$.
We observe that instead of using the same initialization as in (a), taking the Optimal controller as the initial point reduces the number of iterations to converge.
(c) \emph{Unconstrained:} The third learned controller that we design is based on optimizing the objective function of fitting the controller to the sampled data without any consideration of stability. 
%
%\marginJC{WHy don't we simulate (6) for comparison (hopefully, it is not stable).}
%\marginPS{Unfortunately, it's stable.}
%
%maximally sparse, meaning that increasing $\beta$ beyond this value, does not result in any sparsity enhancement. Interestingly, this controller exhibits the same sparsity pattern as the power network and is distributed over $\G$.
%
%\marginJC{But the beta only promotes sparsity, no? This does not actually mean that the controller is distributed, no?}
%
%(c) \emph{Local}: The third controller we design is local, and based on Corollary~\ref{co:local}. It does not take into account the training data and unfortunately, as noted above in (b), simply increasing the penalty parameter $\beta$ in~\eqref{eq:sparse} is not enough to ensure that the resulting controller is local.
%
%\marginJC{How about trying to find a local controller by writing $K$ as $K = [K_1 K_2]$, with $K_1$ and $K_2$ diagonal, and optimize over the entries of the diagonals? SInce $X_3$ is diagonal, then $P$ has also a relatively simple expression.}
%
%Since this controller just ensures stability, it is not optimal for the training data $(\xx^k,\uu^k)$.
%Unfortunately, in our simulations, we observed that simply increasing the penalty parameter $\beta$ in~\eqref{eq:sparse} is not enough to ensure that the resulting controller is local.
%
%\marginJC{This is muddy: in previous sentence, we say "The third controller we design is local", but here we say it is not??}
%
%, and as such, the design of a local controller which is also optimal for the training data is left for future work.

\begin{figure}[tb]
    \centering
    \includegraphics[width=0.47\textwidth]{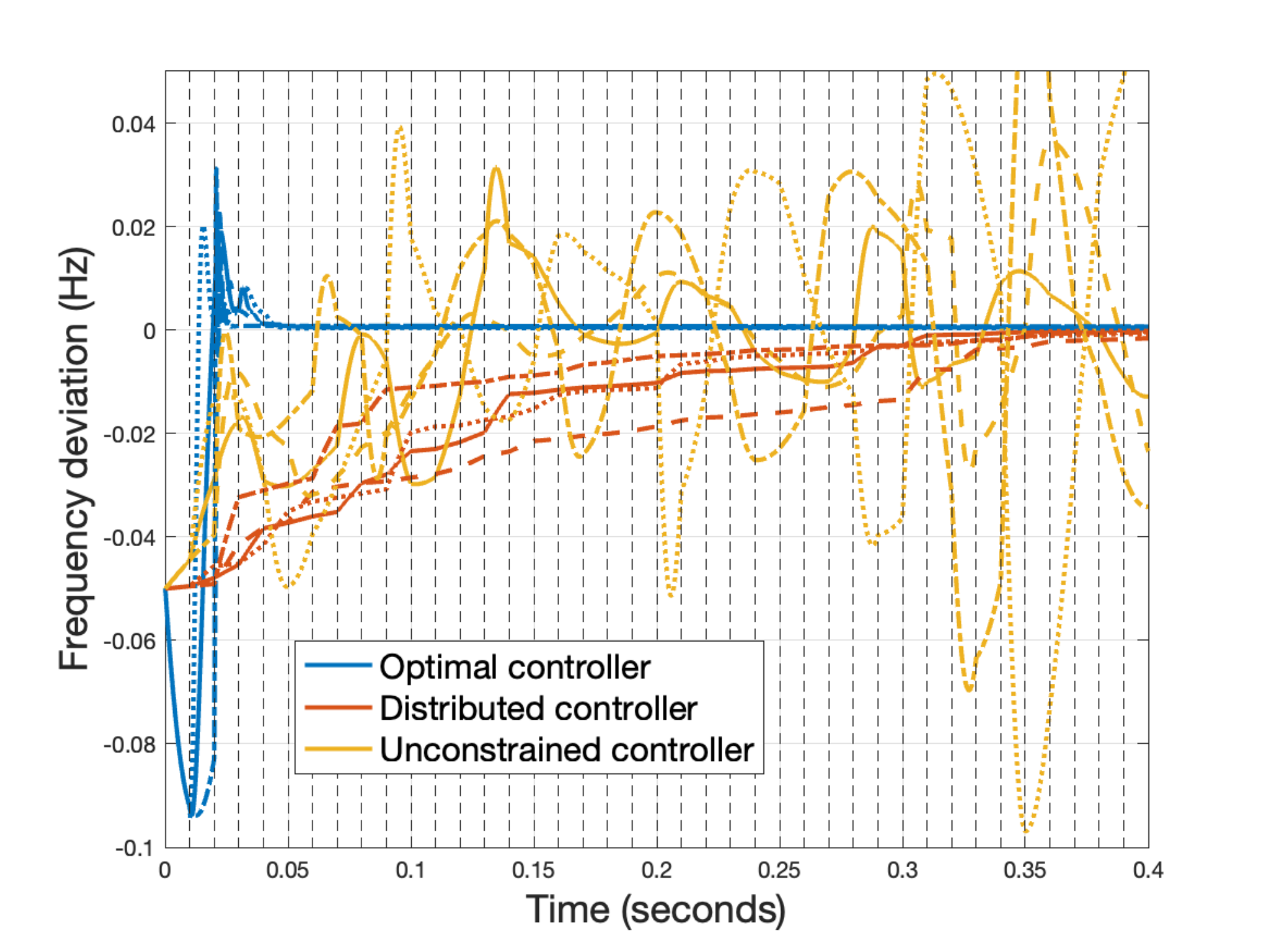}
    \caption{Frequency deviation at node 1 for different switching sequences using the learned controllers. Dashed vertical lines represent the switching instances (every $10^{-2}$ second). Line styles correspond to different switching sequences.}
    \label{fig:evolution}
\end{figure}

To compare the performance of the designed controllers, we display their dynamical response for the same switching sequences. For each simulation, we assume that the system starts in mode 10 (9 seconds of inertia), with an initial frequency deviation of 0.05~Hz at each node, and can switch to any other mode every 0.01 seconds. In Figure~\ref{fig:evolution}, we plot the frequency deviation at node 1 for different switching (inertia) sequences.
Frequency evolution with the unconstrained controller emphasizes the importance of including the stability constraints for the switched system in formulations~\eqref{eq:common_lyap} and~\eqref{eq:sparse}.
It is interesting to note that even though the Optimal controller has an higher overshoot for all the switching sequences, convergence is also faster. To further compare the Optimal and Distributed
controllers, we simulate the dynamics for 1 second in each mode, from an initial frequency deviation of 0.15~Hz at every node.
Table~\ref{tab:metrics} provides the total absolute value of the control input and the total absolute value of frequency deviation for 3 fixed inertia modes $(q=1,5,10)$.
\begin{table}[]
\centering
\caption{Performance metrics for the stable learned controllers under different inertia modes.}
\begin{tabular}{|c|l|c|c|}
\hline
Mode  & Learned Controller & $\int\limits_0^1 \sum\limits_{i=1}^n |u_i(t)| dt$ & $\int\limits_0^1 \sum\limits_{i=1}^n |\omega_i(t)| dt$ \\ \hline 
\hline
\multirow{2}{*}{1}  & Optimal     & $17.38$         & $0.004$       \\ \cline{2-4} 
                    & Distributed & $73.14$        & $0.027$    \\ \hline
\multirow{2}{*}{5}  & Optimal     & $163.66$       & $0.033$       \\ \cline{2-4} 
                    & Distributed & $627.97$      & $0.204$       \\ \hline
\multirow{2}{*}{10} & Optimal     & $726.83$        & $ 0.140$       \\ \cline{2-4} 
                    & Distributed & $1308.10 $       & $ 0.624$   \\ \hline
\end{tabular}
\label{tab:metrics}
\vspace*{-1ex}
\end{table}

As expected, the Optimal controller, which requires state information from all the nodes, 
outperforms the Distributed controller. 
The mean of performance differences taken over the 10 nodes is  62\% for the cumulative control action, and 79\% for the cumulative frequency deviation. This trade-off in performance comes with a saving of 90\% in communication without compromising the system stability.

\section{Conclusions and Future Work}\label{sec:conclusions}
We have presented a framework to synthesize data-driven controllers to regulate the frequency of power networks under time-varying inertia. 
The proposed learning-based design seeks to imitate, under suitable stability constraints, optimal trajectories for different scenarios of changes in inertia generated by finite-horizon LQR formulations. We establish that, regardless of the inertia values, stabilizing learned controllers are guaranteed to exist and are amenable to distributed implementation. Future work will explore the design of efficient algorithms to identify distributed controllers which take optimality with respect to the training data into account and the extension of our approach to nonlinear AC power dynamics. 


\end{document}